\documentclass[runningheads,fleqn]{svmult}
\usepackage{makeidx,epsf,here}   
\usepackage{graphicx}  
\usepackage{subeqnar}  
\usepackage{multicol}  
\usepackage{taphys}    
\makeindex             

\begin{document}
\title*{Semiclassical description of shell effects\\ in finite fermion systems}
\toctitle{Semiclassical description of shell effects
          in finite fermion systems}

\titlerunning{Semiclassical description of shell effects in finite fermion systems}

\author{Matthias Brack}
\authorrunning{Matthias Brack}

\institute{Institute for Theoretical Physics, University of Regensburg,\\
           D-93040 Regensburg, Germany}
 
\maketitle              

\begin{abstract}
\index{abstract}

Since its first appearance in 1971, Gutzwiller's trace formula has been
extended to systems with continuous symmetries, in which not all periodic
orbits are isolated. In order to avoid the divergences occurring in 
connection with symmetry breaking and orbit bifurcations (characteristic 
of systems with mixed classical dynamics), special uniform approximations 
have been developed. We first summarize some of the recent developments 
in this direction. Then we present applications of the extended trace 
formulae to describe prominent gross-shell effects of various finite 
fermion systems (atomic nuclei, metal clusters, and a mesoscopic device) 
in terms of the leading periodic orbits of their suitably modeled 
classical mean-field Hamiltonians.

\end{abstract}

\section{Introduction}

Although Gutzwiller investigated also integrable systems such as the Kepler 
problem in his series of papers \cite{gutz1} beginning in the late 1960s, 
the break through of his semiclassical theory came with the trace formula 
for isolated orbits, published in the last paper, whose 30th anniversary we 
are celebrating this year. This trace formula is most suited for chaotic 
systems in which all periodic orbits are unstable. It has, indeed, launched 
the success of the periodic orbit theory (POT) for the semiclassical 
description of chaotic systems \cite{gubook,pot92,chaos,symp}. Shortly after 
Gutzwiller, Balian and Bloch \cite{bablo} published a trace formula for 
particles in two- and three-dimensional billiards with ideally reflecting 
walls, which may be integrable or non-integrable. The spherical cavity 
investigated by them found a beautiful physical realization in the `supershell' 
structure of metal clusters \cite{supsh} (see also \cite{mbrmp}). Berry and 
Tabor first derived \cite{betab1} a general trace formula for integrable 
systems starting from EBK quantization -- a precursor of their approach may be 
found in \cite{boga} -- and then showed \cite{betab2} that it could also be 
derived from Gutzwiller's semiclassical Green function.  

However, most physical systems are neither integrable nor chaotic, but have 
mixed classical dynamics. When a continuous (dynamical or spatial) symmetry 
is present, the periodic orbits appear in degenerate families and are no 
longer isolated. Starting in 1975, Strutinsky and collaborators \cite{strmag}
generalized Gutzwiller's approach to take into account such symmetries by
performing some of the trace integrations exactly (instead of using the 
stationary-phase approximation). These authors also pioneered the idea of 
employing the POT not for semiclassical quantization but for describing 
{\it gross-shell quantum effects} in mean-field systems in terms of their 
{\it shortest periodic orbits} \cite{strmag,strdos}. For that purpose they 
derived trace formulae not only for the level density, but also for the 
energy shell-correction $\delta E$ (i.e., the oscillating part of the total 
energy of an interacting system; see also \cite{bbook} for details). A more 
general, and mathematically quite elegant, technique of deriving trace 
formulae for systems with continuous symmetries (including integrable systems)
was developed in the early 1990s by Creagh and Littlejohn \cite{crlit}. They 
transformed (part of) the trace integral in the phase-space representation
to an analytical integration over the Haar measure of the symmetry group that 
characterizes the degenerate orbit families. 

With this, trace formulae are available covering all situations from fully 
integrable to fully chaotic systems. The leading amplitudes in a trace 
formula (i.e., those with the lowest order in $\hbar$) come from the most 
degenerate orbit families \cite{strmag,crlit}; less degenerate orbits 
contribute at higher orders in $\hbar$. This leads us to the next problem: 
when the variation of a continuous system parameter (energy, deformation, 
strength of an external field, etc.) causes the breaking or restoring of a 
symmetry, the leading-order amplitudes change discontinuously and diverge 
at the critical points. The same happens when periodic orbits undergo 
bifurcations, which is inevitable in a system with mixed dynamics. Both 
phenomena are closely related and, technically speaking, come from the break 
down of the stationary-phase approximation at the critical points. These 
divergences can be removed by going beyond the first-order saddle-point 
approximation \cite{bablo}, which results in local uniform approximations 
\cite{ozoha}. In order to recover the Gutzwiller amplitudes far from 
the critical points, global uniform approximations must be developed.

In this paper we first review briefly some uniform approximations, without
discussing any technical details, and present two recent examples (Sect.\ 2). 
We then give in Sect.\ 3 a personal account of some applications of the POT 
to the semiclassical description of gross-shell effects in various finite 
fermion systems (nuclei, metal clusters, and a mesoscopic device) in terms 
of the leading periodic orbits of their modeled mean-field Hamiltonians.

\section{Uniform approximations for symmetry breaking and bifurcations}

Uniform approximations can most elegantly be derived using normal forms of the 
action integral in the exponent of the semiclassical Green function in the 
phase-space representation \cite{ozoha}. Tomsovic {\it et al.}\ \cite{tgu} 
derived a general trace formula for the generic breaking of periodic orbit 
families in two-dimensional systems with U(1) symmetry into isolated pairs 
of stable and unstable orbits. Starting from the Berry-Tabor trace formula 
\cite{betab1} in the integrable limit, they generalized the local uniform 
approximation of Ozorio de Almeida and Hannay \cite{ozoha} by means of a 
non-linear coordinate transformation, expanding the Jacobian of this 
transformation and the Van Vleck determinant consistently with the expansion 
of the action integral in the semiclassical Green function, and matching 
the asymptotic Gutzwiller amplitudes and actions of the isolated orbits away 
from the integrable limit. No generally valid trace formula for the breaking 
of higher symmetries have been found so far. 

\begin{figure}[H]
\includegraphics[width=0.92\textwidth]{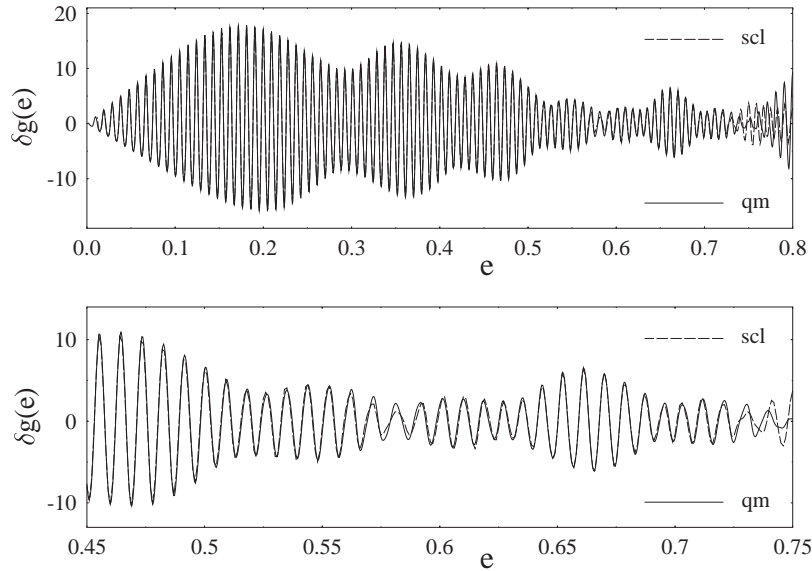}
\caption[]{Level density of the H\'enon-Heiles potential, Gaussian convoluted
over an energy interval $\gamma=0.25\,\hbar\omega$. {\it Solid line:} quantum 
result. {\it Dashed line:} semiclassical result in the uniform approximation 
of \cite{hhun}; only the three shortest primitive orbits and their second 
repetitions are included.}
\label{brafig1}
\end{figure}

Special uniform approximations for the breaking of SU(2) symmetry in 
two-dimensional systems and SO(3) symmetry in a three-dimensional system with 
axial symmetry have been derived by Brack {\it et al.}\ \cite{hhun}. In Fig.\ 
\ref{brafig1} we show their result for the coarse-grained level density of the 
well-known H\'enon-Heiles potential \cite{hh} which has become a paradigm for 
a system with mixed dynamics reaching from near-integrable motion at low 
energy up to nearly chaotic motion at the scaled critical energy (normalized 
to $e=1$) at which the particle can escape over a saddle. An excellent 
agreement between quantum mechanics and semiclassics is reached up to about 
75\% of the critical energy. At low energies, one reaches the SU(2) symmetry 
of the two-dimensional isotropic harmonic oscillator with its regular shell 
structure (frequency $\omega$) and an amplitude linear in the energy (see 
\cite{bbook}). In the region $0.3 < e < 0.75$, the original Gutzwiller trace 
formula for isolated orbits applies \cite{hhbb}, and the uniform approximation 
is seen here to interpolate smoothly down to the integrable limit at $e=0$. 
The break down at $e > 0.75$ is mainly due to bifur- 
\noindent
cations. The straight-line orbit approaching the saddle undergoes an infinite 
cascade of isochronous bifurcations which coalesce at the critical energy 
$e=1$ and pose a series problem to their semiclassical treatment. (See also
\cite{gutzf}, where the self-similarity and Feigenbaum type scaling properties 
of the bifurcated orbits are discussed analytically.)

The most systematic development of uniform trace formulae for all generic
types of bifurcations has been undertaken by Sieber and Schomerus \cite{ssun},
who also used local normal forms and extended them in order to smoothly join 
the asymptotic Gutzwiller amplitudes of the isolated orbits as sketched above. 
Hereby also the analytical continuations of periodic orbits into the complex 
phase space (so-called `ghost orbits' \cite{haake}) contribute in the 
neighborhood of the bifurcations. Interferences of close-lying bifurcations 
(of codimension two) \cite{codim2} and bifurcations of ghost orbits 
\cite{barts} have also been successfully treated with the same technique. 

\begin{figure}[H]
\includegraphics[width=1.0\textwidth]{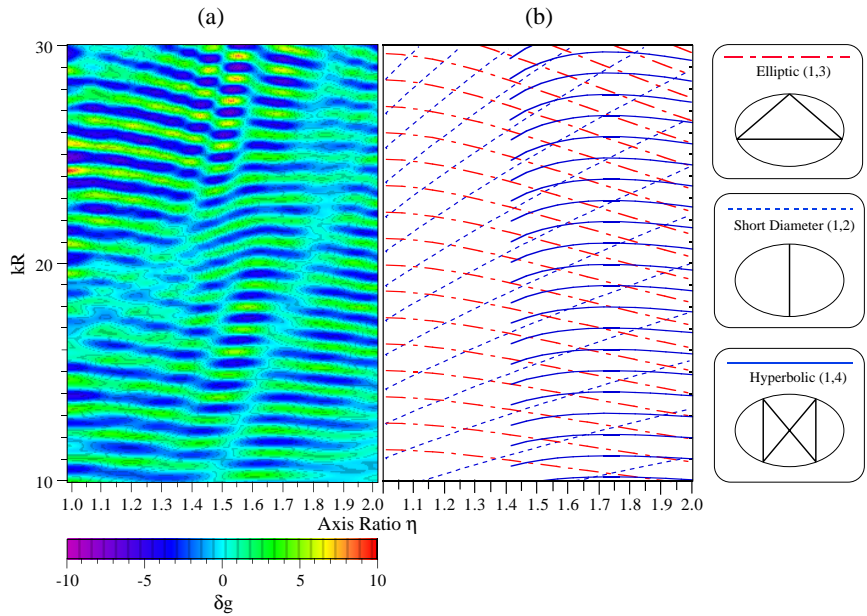}
\caption[]{Contour plot of level density in the ellipse billiard (a) and loci 
of constant actions (b) of its leading periodic orbits. (See text and 
\cite{elli} for details.)}
\label{brafig2}
\end{figure}

In \cite{elli}, an analytical trace formula has been derived for the 
two-dimension\-al ellipse billiard. Although this is an integrable system, it 
exhibits all the complications of mixed systems, including symmetry breaking
and bifurcations. Figure \ref{brafig2} shows in a contour plot (a) its 
coarse-grained oscillating level density $\delta g(E)$ versus wave number $k$
and axis ratio $\eta$. Next to it (b) we see the lines of constant actions
of the shortest periodic orbits illustrated on the right-hand side. The standard 
uniform approximations were not used in \cite{elli}; 
\noindent
the divergences in the spherical limit and at the bifurcations 
of the short diameter orbit (i.e., of its repetitions) could be removed by 
limiting the lowest-order saddle-point integration to the finite limits 
imposed by the classically allowed region. Note that the shell structure 
seen on the left of Fig.\ \ref{brafig2} in some regions of the ($k,\eta$) 
plane is clearly affected by the onset of the new hyperbolic `bow-tie' orbit 
familiy born in a bifurcation at $\eta=\sqrt{2}$. An extension of this study 
to the three-dimensional spheroidal cavity is in progress \cite{sphero}.

A simple, but efficient way to avoid the difficulties connected with symmetry 
breaking and bifurcations, at least in some situations, is to use a 
perturbative trace formula developed by Creagh \cite{crpert}. In this 
approach which, of course, can only be used for sufficiently small deviations 
from an integrable limit, the effect of a non-integrable perturbation is only 
taken into account in the actions of the periodic orbits; the stability 
amplitudes and Maslov indices of the integrable system are kept unchanged. 
(A similar approach was used also in \cite{pertb}.) This results in the 
modification of the integrable-limit trace formula merely by a modulation 
factor which contains the average of the lowest non-vanishing perturbation of 
the action over each unperturbed orbit family, and which often can be calculated 
analytically \cite{crpert,pmei,hhpert}. An application of the perturbative trace
formula is given in Sect.\ \ref{nucdef} below.

\section{Applications to shell structure in finite fermion systems}

\subsection{Ground-state deformations of nuclei and metal clusters}
\label{nucdef}

\begin{figure}[H]
\includegraphics[width=.76\textwidth]{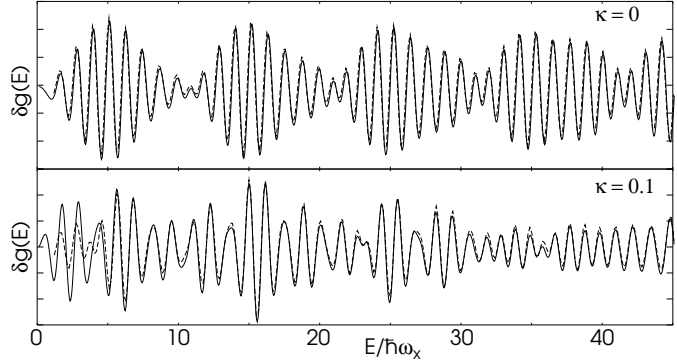}
\caption[]{Coarse-grained level density of a three-dimensional harmonic
oscillator with frequencies $\omega_x=1$, $\omega_y=1.12128$, $\omega_z=
1.25727$ (Gaussian averaging range $\gamma=0.2\,\hbar\omega_x$), both 
without (top) and including a spin-orbit interaction (bottom). {\it Solid 
lines:} quantum-mechanical, {\it dotted lines:} semiclassical results (see 
\cite{luso} for details).}
\label{brafig4}
\end{figure}

An early application of the POT to explain the systematics of ground-state 
deformations of atomic nuclei was given by Strutinsky {\it et al.}\ 
\cite{strdos}. In contour plots of the quantum-mechanically calculated energy 
shell-correction $\delta E$ versus nucleon numbers $N$ and deformation 
parameter $\eta$, the correct slopes of the minimum valleys are reproduced by 
the systematics predicted from the leading periodic orbits of a spheroidal 
cavity. A more complete study in the same model was given later by Frisk 
\cite{frisk}, and a detailed Fourier analysis of its quantum spectrum was 
performed by Arita {\it et al.}\ \cite{matsu}, who also discussed the role of 
orbit bifurcations (without, however, developing the appropriate trace 
formulae). All these authors have neglected the spin-orbit interaction; 
although it is known to modify the shell structure in nuclei (cf.\ Fig.\ 
\ref{brafig4}), its effect could be simulated by a simple renormalization of 
the Fermi energy (see also Sect.\ \ref{fission} below). 

\begin{figure}[H]
\includegraphics[width=.63\textwidth]{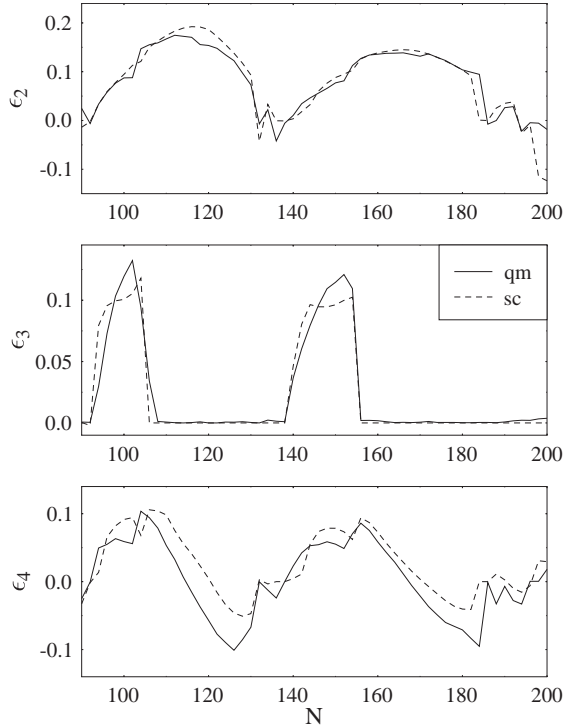}
\caption[]{Ground-state quadrupole ($\epsilon_2$), octopole ($\epsilon_3$), 
and hexadecapole ($\epsilon_4$) deformations of sodium clusters versus 
number $N$ of valence electrons, using axially deformed cavities as their 
mean potential. {\it Solid lines:} quantum results. {\it Dashed lines:} 
semiclassical results using the perturbative trace formula \cite{crpert,pmei}. 
(From \cite{pashk}.)}
\label{brafig3}
\end{figure}

Luckily, the ground-state deformations of not too light nuclei and metal 
clusters are sufficiently small so that the perturbative trace formula of 
Creagh \cite{crpert} may be applied successfully. In Fig.\ \ref{brafig3} we 
show a recent comparison of the lowest multipole deformations of sodium 
clusters, calculated \cite{pashk} both quantum-mechanically and 
semiclassically with the perturbative trace formula using the modulation 
factors derived in \cite{pmei}. The total energy of each cluster with fixed 
particle number $N$ was minimized with repect to all three 
\noindent
deformation parameters. Their equilibrium values obtained in the two ways 
are seen to agree almost quantitatively, which demonstrates the usefulness 
of the perturbative semiclassical approach. Note that the spin-orbit 
interaction plays a negligible role in sodium clusters \cite{mbrmp}.

As a first step towards the inclusion of the spin-orbit interaction in the 
semiclassical trace formula for nuclei, we show in Fig.\ \ref{brafig4} the 
level density obtained recently \cite{luso} for a three-dimensional deformed 
harmonic oscillator which is a realistic model for light nuclei. Hereby the 
approach of Littlejohn and Flynn \cite{lifly} was employed in the same 
heuristic way as in \cite{frgu}. Instead of giving more details of this 
approach, we refer to a more rigorous semiclassical theory including spin 
degrees of freedom \cite{boke}. We see in Fig.\ \ref{brafig4} that the 
spin-orbit interaction does drastically change the gross-shell structure, and 
that it can be described semiclassically, indeed. In the present example the 
still unsolved mode-conversion problem (occurring along manifolds in phase
space where the spin-orbit interaction locally is zero) did not arise. Some 
steps towards its solution in a two-dimensional system are in progress 
\cite{modcon}.

\subsection{Mass asymmetry in nuclear fission}
\label{fission}

\begin{figure}[H]
\includegraphics[width=.75\textwidth]{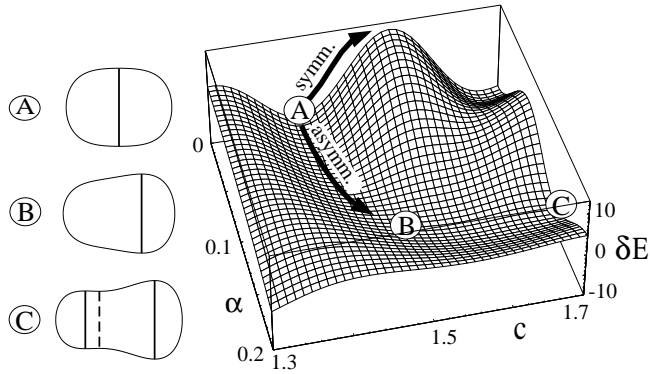}
\caption[]{Fission barrier of a heavy nucleus. (From \cite{fiss}; 
see text for details.)}
\label{brafig5}
\end{figure}

Another example for the contribution of periodic orbits to a prominent quantum 
shell effect in a complex interacting fermion system is the asymmetry in the 
fission of heavy nuclei, which results in an asymmetric distribution of the 
fission fragments. This asymmetry, which sets in already during the passage
over the saddle in the deformation energy space, has long been taken as a
prime example of a quantum phenomenon that could not be explained classically,
e.g., in terms of the liquid-drop model. (For a detailed presentation of the 
role of shell effects in nuclear fission see, e.g., the review \cite{fuhil}.) 
The POT, however, allows to understand this effect semiclassically, using only 
very few periodic orbits \cite{fiss,tsuk}. Figure \ref{brafig5} shows a 
pespective view of the deformation energy of a typical heavy nucleus, plotted 
versus elongation parameter $c$ and asymmetry parameter $\alpha$. It was 
calculated in \cite{fiss} using the trace formula for 
\noindent
the shell-correction energy $\delta E$ of particles in an axially deformed 
cavity with the shapes defined in \cite{fuhil}. A coarse-graining simulating  
the pairing interaction was used. The lowest adiabatic path to fission, 
determined by the stationarity of the actions of the leading orbits (cf.\ 
\cite{strdos}), leads from the isomer minimum (point A) over a saddle with 
asymmetric shapes ($\alpha>0$, points B and C). Imposing symmetry ($\alpha=0$) 
would lead over a barrier at appreciably higher energy. This is exactly the 
topology of the fission barrier obtained in the old quantum-mechanical 
calculations with realistic nuclear shell-model potentials \cite{fuhil}. Only 
few periodic orbits need to be included to obtain the semiclassical result. 
They lie in planes perpendicular to the symmetry ($z$) axis, as illustrated 
to the left of Fig.\ \ref{brafig5} by the perpendicular lines (solid for 
stable and dashed for unstable orbits) drawn into the shapes corresponding to 
the three points in deformation space. These orbits are just the polygons 
inscribed into the circular cross sections of the cavity with those planes; 
their stability amplitudes were given by Balian and Bloch \cite{bablo}. 
A uniform approximation was used in \cite{fiss} to handle the bifurcation 
happening when the cavity starts to neck in and the plane containing the 
shortest orbits splits into three planes (existing, e.g., at point C). As 
shown in \cite{tsuk}, only primitive orbits with up to $\sim 5$ reflections 
were needed in each plane to obtain a converged result; the two shortest 
orbits (diameter and triangle) were, in fact, sufficient to obtain the correct 
topology of the asymmetric fission barrier. As in \cite{strdos,frisk}, the 
spin-orbit interaction was neglected; instead, the Fermi energy was adjusted 
such that the isomer minimum appeared at the correct deformation.

\begin{figure}[H]
\leavevmode
\epsfysize=5cm
\epsfbox[-66 159 664 683]{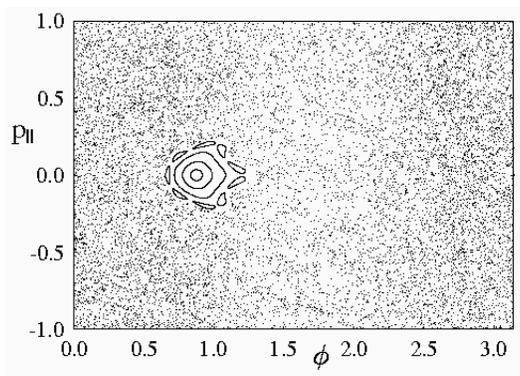}
\caption[]{Poincar\'e surface of section of classical trajectories with $L_z=0$ 
at deformation of asymmetric saddle (B). At each mapping (reflection off the 
boundary), $\phi$ is the polar angle and $p_{||}$ the momentum component 
parallel to the tangent plane.}
\label{poinc}
\end{figure}

It is interesting to notice that the classical motion in the cavities with 
shapes occurring around the fission barrier is quite chaotic, as discussed in 
more detail in \cite{nobfis}. In Fig.\ \ref{poinc} we show a Poincar\'e surface 
of section, taken at the asymmetric saddle (near point B) for a number of 
trajectories starting from random initial conditions with angular momentum
$L_z=0$. It reveals us
\noindent
that this part of the phase space is, indeed, more than 95\% chaotic. Only a 
small regular island surrounds the fixed point corresponding to the diameter 
orbit. It is this small regular island, embedded in a chaotic phase space, 
that hosts the periodic orbit which is chief responsible for the shell effect 
driving the nucleus to asymmetric shapes. In the microscopic description, the 
quantum-mechanical states responsible for this shell effect are a few 
`diabatic' states whose eigenenergies depend strongly on the asymmetry 
parameter, causing the energy gain in going from the symmetric to the 
asymmetric saddle, whereas most other states are insensitive to it (cf.\ \cite{gumn}). 
The diabatic quantum states in the present cavity model were shown in 
\cite{nobfis} to have their probability maxima exactly in the planes 
containing the shortest periodic orbits. Furthermore, an approximate EBK 
quantization of the classical motion near those planes reproduces the 
eigenergies of the diabatic quantum states almost quantitatively \cite{nobfis}, 
thus establishing a nice quantum-to-classical correspondence in a highly 
nonlinear complex system.

\subsection{Mesoscopic systems}

\begin{figure}[H]
\includegraphics[width=.9\textwidth]{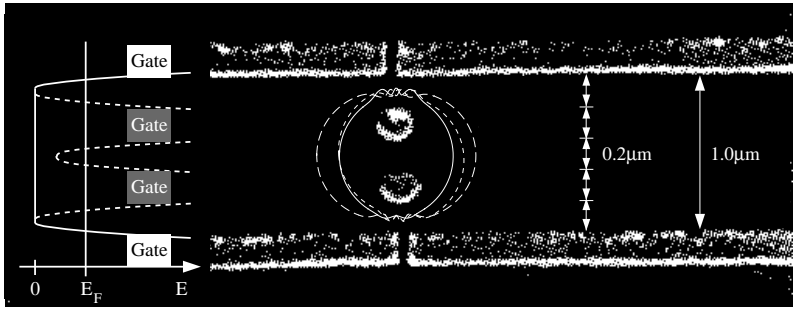}
\caption[]{Mesoscopic channel with two antidots (from \cite{chan}). {\it Left:}
sketch of the model potential confining the electrons, and position of the Fermi 
energy $E_F$.}
\label{brafig7}
\end{figure}

\begin{figure}[H]
\includegraphics[width=.6\textwidth]{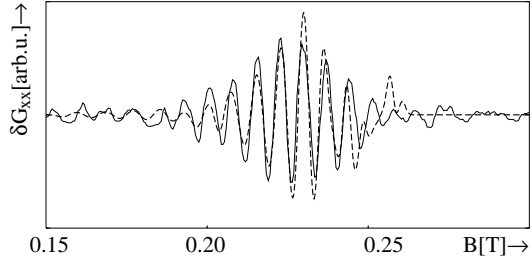}
\caption[]{Comparison of experimental conductance oscillations (solid line) 
and semiclassical result (dashed line) with optimized parameters of the
model potential \cite{jphd}.}
\label{brafig8}
\end{figure}

We finally turn to a mesoscopic arrangement in which a two-dimensional electron 
gas is confined laterally to a channel of width $\sim 1.0\,\mu$m. Two antidots 
represent obstacles to the electric current through the channel; the effective 
radius of these antidots can be regulated by an applied gate voltage $V_g$. 
Figure \ref{brafig7} shows an SEM photograph of the experimental gate structure 
\cite{goul}. The longitudinal conductance $G_{xx}$ along the channel was 
measured for various strengths of a perpendicular magnetic field $B$ and 
various gate voltages $V_g$ \cite{goul,kirc}. A commensurability minimum in the 
average conductance was observed near those values of $B$ for which a cyclotron 
orbit fits around the antidots. Small observed oscillations around the average 
part of $G_{xx}$ could be interpreted semiclassically \cite{chan,jphd} by the 
interferences of the leading periodic orbits (a few of which are shown in Fig.\ 
\ref{brafig7} by solid and dashed white lines). In Fig.\ \ref{brafig8} we compare 
the experimental oscillations $\delta G_{xx}$ with the result of the 
calculation \cite{jphd} in which the semiclassical Kubo formula \cite{kubosc}
was used. 
\noindent

\noindent

\begin{figure}[H]
\includegraphics[width=.8\textwidth]{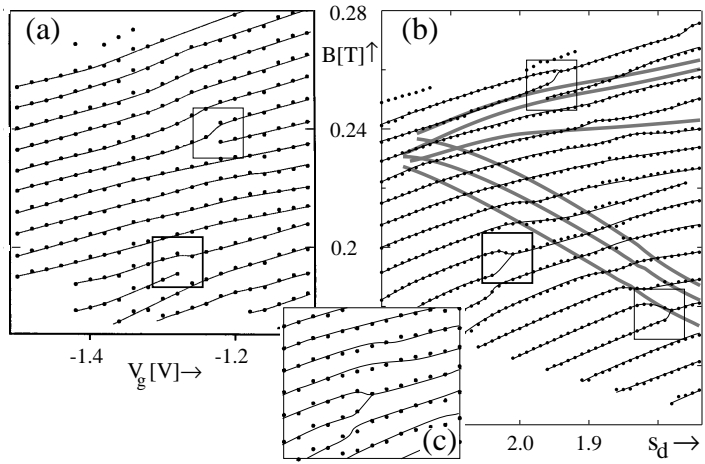}
\caption[]{Maximum positions of $\delta G_{xx}$ versus $B$ (vertical axes) 
and $V_g$ (horizontal axes). (a) Experimental values \cite{kirc}. (b) 
Semiclassical results \cite{chan}; $s_d$ is the antidot radius regulated by 
$V_g$ (approximately one has $s_d\propto V_g$); the gray-shaded lines 
correspond to the loci of bifurcations of some leading orbit families. 
(c) Behaviour near a dislocation (dots: experiment; lines: semiclassical 
results). (From \cite{chan}.)}
\label{brafig9}
\end{figure}

An interesting phenomenon is observed when varying both the magnetic field $B$ 
and the gate voltage $V_g$ and plotting the loci of the oscillation maxima in 
$\delta G_{xx}$. These arrange themselves, as seen in Fig.\ \ref{brafig9} (a), 
along smooth lines whose slopes are well understood in terms of the $B$ and 
$V_g$ dependence of the actions of the leading periodic orbits. However, some 
characteristic dislocations occur at apparently random places in the $(B,V_g)$ 
plane, as emphasized by the boxes. In the semiclassical analysis, they 
originate from successive bifurcations of periodic orbits: the different orbit 
generations lead to different slopes in Fig.\ \ref{brafig9} (b), and these do 
not match near the loci in the $(B,V_g$) plane (shown for some leading orbits 
by gray-shaded thick lines) along which the bifurcations occur. Although the 
theory does not fit the experiment globally 
\noindent
(at least 10 different orbit families contribute), the local agreement near 
the dislocations is excellent; see the box in Fig.\ \ref{brafig9} (c). A
quantum-mechanical calculation \cite{kirc} qualitatively reproduced the 
dislocations, too. But the physical understanding of their origin required 
the semiclassical analysis in terms of periodic orbits. As we see, even the 
orbit bifurcations have experimentally observable consequences!

\bigskip
\bigskip

I am grateful to all my students and collaborators whose work has been
presented here, and to the Deutsche Forschungsgemeinschaft for partial
financial support.

\end{document}